\documentclass[12pt]{article}
\usepackage{amsmath}
\usepackage{tikz}
\usepackage{graphicx}  
\usepackage{adjustbox} 
\usetikzlibrary{positioning, calc}
\usepackage[letterpaper, margin=1in]{geometry}
\usepackage[linesnumbered,ruled,vlined]{algorithm2e}

\usepackage{amsmath,amsthm,amssymb,amsfonts,verbatim,}
\usepackage[colorlinks,
            linkcolor=blue,
            anchorcolor=blue,
            citecolor=blue
            ]{hyperref}
\usepackage{array}
\usepackage{multirow}

\usepackage{color}
\usepackage{booktabs}

\usepackage{setspace}

\title{Efficient construction of subspace design codes over constant alphabet}

\author{
Shu Liu \thanks{\sf email: shuliu@uestc.edu.cn} 
\\
{\small National Key Laboratory of Wireless Communications}\\{\small
University of Electronic Science and Technology of China.}}

\date{}

\newtheorem{lemma}{Lemma}[section]
\newtheorem{theorem}[lemma]{Theorem}
\newtheorem{cor}[lemma]{Corollary}

\newtheorem{ex}[lemma]{Example}
\newtheorem{rem}[lemma]{Remark}
\newtheorem{defn}{Definition}

\newtheorem{main}{Main Theorem}

\theoremstyle{remark}

\renewcommand{\epsilon}{\varepsilon}
\renewcommand{\le}{\leqslant}
\renewcommand{\ge}{\geqslant}

\def\mO{{\mathcal O}}

\newcommand{\vnote}[1]{}

\def\ZZ{\mathbb{Z}}
\def\PP{\mathbb{P}}

\def\F{\mathbb{F}}
\def \mC {\mathcal{C}}
\def \mA {\mathcal{A}}

\def \mA {\mathcal{A}}

\def \mC {\mathcal{C}}

\def \mL {\mathcal{L}}

\def\cL{{\mathcal L}}

\def\Pin{{P_{\infty}}}

\def \g {{\sf g}}

\newcommand{\Ga}{\alpha}
\newcommand{\Gb}{\beta}
     
\newcommand{\Gd}{\delta}     
\newcommand{\Ge}{\epsilon}

\newcommand{\Gl}{\lambda}    
\newcommand{\Go}{\omega}     \newcommand{\GO}{\Omega}
\newcommand{\Gs}{\sigma}

\def\AG{algebraic geometry
 }

\def \bc {{\bf c}}
\def \bi {{\bf i}}

\def \bu {{\bf u}}

\def \bo {{\bf 0}}
\def\Supp {{\rm Supp }}

\def\Aut {{\rm Aut }}

\def\Fm{{\F_{q^m}}}

\begin{document}
\maketitle

\begin{abstract} 
Subspace-design codes have recently emerged as a central unifying framework in coding theory, serving as a bridge between probabilistic existence results and explicit, algebraically structured constructions. The two fundamental code families are folded Reed–Solomon (FRS) codes and univariate multiplicity codes, both possessing the subspace-design property. These codes attain the list-decoding capacity with polynomial-sized lists, yet they require an alphabet size scaling with the code length, which constrains their practical efficiency. A recent breakthrough by Goyal, Guruswami, and Hsieh yields the first explicit construction of subspace-design codes over constant-size alphabets via the Alon–Edmonds–Luby (AEL) expander-based framework. This settles a major open problem and shows that the structural guarantees of subspace designs can be attained without large alphabets.

Despite recent progress on explicit constructions, the problem of explicitly constructing subspace-design codes-particularly those that admit efficient, low-complexity encoding-remains a central challenge in the field. In this paper, we present an explicit construction of $r$-dimensional subspace-design codes over a constant-sized alphabet of size $\exp(40r)$. In comparison, the $r$-dimensional subspace-design codes proposed by Goyal et al. (FOCS 2026) require an alphabet size of $\exp(\exp(O(r^2)))$.  In addition, our construction is computationally efficient, with the encoding procedure requiring only $\Tilde{O}(rn^2)$ field operations. Our framework is built upon a modified variant of algebraic geometry (AG) codes, where rational evaluation points are substituted with higher-degree places. This methodology yields a valid additive code. To validate the favorable parameter guarantees of the constructed code, we leverage Moore matrices to analyze determinants with abundant zero entries.  Furthermore, we generalize univariate multiplicity codes, which are traditionally constructed from Reed–Solomon codes, to the setting of algebraic geometry codes. We establish that these AG-multiplicity codes over non-constant alphabets also achieve optimal parameters. 
\end{abstract}

\section{Introduction}
 Originally introduced by Guruswami and Xing \cite{GX13} to construct the first positive-rate rank-metric codes list-decodable beyond half the distance, the concept has since evolved into a powerful higher-dimensional generalization of the minimum distance property. Informally, a subspace-design code is an additive code with the property that every low-dimensional linear subspace of codewords has its zero coordinates spread evenly across the code's positions. This structural guarantee—originally established for folded Reed–Solomon (FRS) and univariate multiplicity codes—has proven to be the key to achieving optimal performance across multiple fundamental metrics simultaneously. Explicitly constructing subspace-design codes-particularly those admitting efficient, low-complexity encoding-remains a central challenge in this area despite recent work on explicit construction by Goyal, Guruswami, and Hsieh \cite{GGH26}. 
 
\subsection{Literature review}
Introduced by Guruswami and Xing \cite{GX13}  in 2013, subspace designs were originally developed to give the first randomized construction of optimal-rate list-decodable codes over constant-sized alphabets with sub-logarithmic list sizes. The central idea is a higher-dimensional generalization of the minimum distance property: an additive code is a subspace-design code if every low-dimensional linear subspace of codewords has its zero coordinates spread evenly across the code's positions.

The foundational explicit constructions came from Guruswami and Kopparty \cite{GK16}, who showed that folded Reed-Solomon codes and univariate multiplicity codes satisfy the subspace-design property with near-optimal parameters. This connection is particularly striking because folded Reed-Solomon codes were themselves the very list-decodable codes that motivated subspace designs in the first place. The parameters of these constructions were later tightened by Brakensiek, Chen, Dhar, and Zhang \cite{BCDZ25b}, who also proved an impossibility result: better subspace designs do not exist over algebraically closed fields.

A major conceptual breakthrough was the establishment of a local equivalence between random linear codes and subspace-design codes. Brakensiek, Chen, Dhar, and Zhang \cite{BCDZ25a}  showed that any local property—such as list decodability or list recoverability—that holds for random linear codes applies to all subspace-design codes, and conversely. This framework provided the first explicit construction of folded linear codes that simultaneously attain all local properties of random linear codes.

The practical significance of subspace-design codes was further demonstrated through proximity gaps. Goyal and Guruswami \cite{GG25} proved that subspace-design codes achieve optimal proximity gaps approaching the information-theoretic bound of 
$1-R$. Their proof introduced the notion of curve decodability and established mutual correlated agreement, a stronger property that implies proximity gaps. This resolved a major open problem, with direct implications for SNARK efficiency.

A recent milestone addressed the alphabet-size bottleneck. Explicit algebraic codes such as folded Reed–Solomon and multiplicity codes require alphabet sizes that grow polynomially in the block length. Goyal, Guruswami, and Hsieh \cite{GGH26} resolved the open question posed by Brakensiek et al. \cite{BCDZ25b}, by providing explicit constructions of subspace-design codes over constant-sized alphabets via the expander-based Alon–Edmonds–Luby (AEL) framework. This generalizes and unifies recent work demonstrating that such codes inherit the local properties of random linear codes.

\subsection{Our results}
In this paper, we show that algebraic geometry codes, evaluated at places of high degree, form subspace design codes  (see Definition 1 in Subsection 2.4) with good parameters. Specifically, we establish the following result.

\begin{main}
  Let $q$ be a prime power. For any integer  $r \ge 1$ and real $R\in(0,1)$, there is an $\mathbb{F}_q$-linear $(\tau,r)$-subspace design code of rate $R$ over an alphabet of size $O(\exp(40r))$, with length tending to $\infty$, and $\tau = R + \Ge(q,r,R)$, where
\[
\Ge(q,r,R)=\frac1r\log_q \left[1+(q^r-1)\left(R+\frac{1}{q^{20r}-1}\right)\right]-R.
\]
Furthermore, encoding of the code requires $\Tilde{O}(rn^2)$ operations over $\mathbb{F}_q$. 
\end{main}

\begin{rem}{\rm \begin{itemize}
    \item[(i)]
It can be verified that   $\frac1r\log_q \left[1+(q^r-1)\left(R+\frac{1}{q^{20r}-1}\right)\right]>R$ for any $r\ge 1$  and a real $R\in(0,1)$.
\item[(ii)] Compared with the subspace-design code given in ~\cite{GGH26},
our code does not attain the optimal bound \(\tau \approx R\). Nevertheless, its alphabet size is \(\exp(40r)\), whereas the subspace-design code from~\cite{GGH26} has alphabet size \(\exp(\exp(O(r^2)))\). Moreover, encoding our code requires only $\Tilde{O}(r n^2)$ operations over \(\mathbb{F}_q\) via pre-computation of a generator matrix of the code. In fact, for a large class of algebraic geometry codes, we can have quasi-linear time encoding (see \cite{LLMWX25}). For these algebraic geometry codes, we have encoding complexity $\Tilde{O}(r n)$
\item[(iii)] In \cite{GXY18}, Guruswami et al provided a construction of subspace design codes based on function fields and Moore matrices. As a large cyclic automorphism group of function field is required for their construction, subspace design codes given in \cite{GXY18} do not yield constant alphabet size. 
\end{itemize}
}\end{rem}
Our second main result, which pertains to the proximity gap (Definition 2, Subsection 2.4), follows directly as a corollary of Main Theorem 1.
\begin{main}
For any integers \(r, m \in \mathbb{N}\) and real $R\in(0,1)$, there exists a subspace design code  over $\F_q^{40r}$ and has rate $R$ and proximity gap
\[
\left( 1 -R- \Ge(q,r,R) - \frac{3}{r}, \frac{r (m(1 - R- \Ge(q,r,R)) + r^2)}{2q}, \frac{1}{m} \right).
\]
\end{main}
\begin{rem}{\rm  \begin{itemize}
    \item[(i)]
For $R=0$, the proximity gap $$1 -R- \Ge(q,r,R) - \frac{3}{r}\approx 1-\frac1r\times \frac1{\ln q}\times \frac1{q^{19r}}-\frac3r\rightarrow 1$$ for large $r$. This implies that
the proximity gap given in Main Theorem 2 exceeds the unique decoding radius $\frac{1-R}2$ for small rate $R$ as both the functions $1 -R- \Ge(q,r,R)$ and $\frac{1-R}2$ are continuous in variable $R$.  
\item[(ii)] To have $ \frac{r (m(1 - R- \Ge(q,r,R)) + r^2)}{2q}<1$, we can take $q=\Theta(rm+r^2)$, i.e., the alphabet size $q^{40r}=\exp(O(r\log(rm)))$ of our code is independent of the code length.
\end{itemize}
}\end{rem}

By Lemma \ref{lem:2.5}, we obtain the following result on list recovery.
\begin{main}
For any integers \(\ell \in \mathbb{N}\) and real $\rho, R\in(0,1)$, set $r=\left\lceil\frac{\ell}{\rho}\right\rceil$. Then there exists a subspace design code  over $\F_q^{40r}$ that
is
$
 \left(1-R-\Ge\left(q,r,R\right)
       -\rho,\ \ell,\ L\right)$-list recoverable,
where
\[
 L\leq
 \left(
   \frac{\ell}{
   \Ge\left(q,r,R\right)+R+\rho}
 \right)^{
 \left(
   \Ge\left(q,r,R\right)+R+\rho
 \right)/\rho}.
\]
\end{main}
\begin{rem}{\rm 
For $R=0$ and fixed $\ell$, the list recovery radius
$
1 - R - \operatorname{\Ge}(q,r,R) - \rho
\approx 1-\frac{1}{r}\times \frac{1}{\ln q}\times \frac{1}{q^{19r}}-\frac{\ell}{r}
\to 1
$
for large $r$. This implies that the list recovery radius given in Main Theorem~3 exceeds the unique decoding radius. Furthermore, our code has alphabet size $\exp(40\ell/\rho)$, while the code given in~\cite{GGH26} has alphabet size $\exp(\exp(O(\ell^2/\rho^2)))$. Note that the code given in~\cite{GGH26} achieves the optimal list recovery radius $1-R-\rho$.
}\end{rem}

\subsection{Our construction and techniques}

Algebraic geometry (AG) codes are a natural generalization of Reed--Solomon codes. 
Indeed, Reed--Solomon codes correspond to the special case of AG codes based on curves 
of genus $0$, namely the projective line. Two families of optimal subspace-design codes 
have been explicitly constructed from variants of Reed--Solomon codes: folded Reed-Solomon codes 
and univariate multiplicity codes. To obtain explicit constructions of optimal subspace-design codes over constant-sized alphabets, it is therefore natural to extend these two frameworks to the setting of algebraic geometry codes.

\subsubsection{Folded algebraic geometry codes}

Folded algebraic geometry codes have been studied in several works in the literature 
(e.g., see \cite{GX14,GX16,GXY18}). We briefly recall their construction. Let $F/\mathbb{F}_q$ be a 
function field equipped with an automorphism $\sigma \in \operatorname{Aut}(F/\mathbb{F}_q)$. 
Choose $n$ pairwise distinct rational places $P_1,\dots,P_n$ such that the places
\[
P_1, P_1^\sigma, \dots, P_1^{\sigma^{s-1}}, \;
P_2, P_2^\sigma, \dots, P_2^{\sigma^{s-1}}, \;
\dots, \;
P_n, P_n^\sigma, \dots, P_n^{\sigma^{s-1}}
\]
are all pairwise distinct for some integer $s\ge 1$. Let $G$ be a divisor of $F$ that is invariant under $\sigma$, 
i.e., $G^\sigma = G$. A folded AG code can then be defined (see \cite{GX14} for details). As in the case of folded Reed--Solomon codes, folded algebraic geometry (AG) codes also form a family of subspace-design codes that achieve optimal parameters (see \cite{GXY18} for details). The main obstacle to obtaining subspace-design codes over constant-sized alphabets from folded AG codes is that such a construction requires the existence of an automorphism $\sigma \in \operatorname{Aut}(F/\mathbb{F}_q)$ whose order is proportional to the code length. However, it was shown in \cite{MX19} that the order of any automorphism of a function field is asymptotically much smaller than the genus as the genus tends to infinity. Consequently, folded algebraic geometry codes do not yield subspace-design codes over constant-sized alphabets. This limitation highlights a fundamental trade-off between the algebraic structure required for folding and the asymptotic constraints imposed by the automorphism groups of function fields.

\subsubsection{Multiplicity algebraic geometry codes and our techniques}

To the best of our knowledge, univariate multiplicity codes have not yet been generalized to algebraic geometry (AG) codes in the existing literature. Unlike the rational function field $\mathbb{F}_q(x)$, where derivatives are naturally taken with respect to the variable $x$, the situation for an arbitrary function field $F/\mathbb{F}_q$ is more subtle: the space of derivations forms a one-dimensional module over $F$ (see \cite[Lemma~4.1.6]{St09}). Thus, a suitable notion of derivative must be chosen in order to define multiplicity AG codes.

\subsection*{Derivatives and the Multiplicity Criterion}

We first recall how derivatives are used in the definition of univariate multiplicity codes. Let $\alpha_1,\dots,\alpha_n \in \mathbb{F}_q$ be pairwise distinct, and let $P_i$ denote the zero place of $x - \alpha_i$. Let $P_\infty$ be the unique pole of $x$. For a function $f \in \mathcal{L}(m P_\infty)$—that is, a polynomial $f(x)$ of degree at most $m$—the encoding is given by
\[
\begin{pmatrix}
    f(P_1) & f(P_2) & \cdots & f(P_n) \\
    f^{(1)}(P_1) & f^{(1)}(P_2) & \cdots & f^{(1)}(P_n) \\
    \vdots & \vdots & \ddots & \vdots \\
    f^{(s-1)}(P_1) & f^{(s-1)}(P_2) & \cdots & f^{(s-1)}(P_n)
\end{pmatrix},
\]
where $f^{(i)}$ denotes the $i$-th derivative with respect to $x$. A key property is that $f$ has a zero of multiplicity at least $r$ at $P_i$ if and only if
\[
f(P_i) = f^{(1)}(P_i) = \cdots = f^{(r-1)}(P_i) = 0.
\]

To generalize this construction to AG codes, one must find a function $t \in F$ such that the derivative with respect to $t$ satisfies an analogous multiplicity criterion: $f$ has multiplicity at least $r$ at $P_i$ if and only if $
f(P_i) = f^{(1)}(P_i) = \cdots = f^{(r-1)}(P_i) = 0.$

This can be achieved as follows. Let $\Pin, P_1, \dots, P_n$ be pairwise distinct rational places of $F$ and let $a_1,\dots,a_n \in \mathbb{F}_q$. By the weak approximation theorem~\cite[Theorem 1.3.1]{St09}, there exists a function $t \in F$ satisfying
\[
\nu_{\Pin}(t) = -1 \qquad\text{and}\qquad \nu_{P_i}(t - a_i) = 1 \quad \text{for all } i = 1,\dots,n.
\]
In the special case of the rational function field $\mathbb{F}_q(x)$, the parameter $t$ is simply $x$, with
\[
\nu_{P_\infty}(t) = -1 \qquad\text{and}\qquad \nu_{P_i}(t - a_i) = 1 \quad \text{for all } i = 1,\dots,n.
\]

This observation forms the basis of our generalization of multiplicity codes to arbitrary algebraic geometry codes.
However, carrying out this generalization requires overcoming two main obstacles, which we discuss in turn below.

\subsection*{First Obstacle: Derivative Existence}

The above construction provides a local parameter $t$ that behaves like the variable $x$ in the rational case. However, the derivative $\frac{d}{dt}$ is not a global derivation on $F$; it only acts on the completion at each place. Nevertheless, for the purpose of evaluating derivatives at a fixed place $P_i$, the local derivative with respect to $t$ is well-defined and satisfies the multiplicity criterion locally. Thus, the first obstacle—finding a suitable derivative—is overcome by using local parameters obtained from weak approximation.

\subsection*{Second Obstacle: Wronskian Nonvanishing}

A second, more serious obstacle is establishing that the Wronskian determinant
\[
W(f_1,\dots,f_r) = \det\!\big( f_j^{(i-1)} \big)_{1 \le i,j \le r}
\]
is nonzero if and only if the functions $f_1,\dots,f_r$ are $\mathbb{F}_q$-linearly independent. In the classical Reed--Solomon setting, this holds as long as the degree $m$ is less than the characteristic $p$ of $\mathbb{F}_q$. Unfortunately, the same condition is required in the case of multiplicity algebraic geometry codes. That is, we must restrict to a subspace of $\mathcal{L}(m \Pin)$ with $m < 2g+p-1$ to ensure the Wronskian criterion. Since $p$ is fixed for a given finite field, this imposes a severe constraint on the code rate and prevents achieving optimal parameters over constant-sized alphabets.

\subsubsection{Generalized algebraic geometry codes}
Given the limitations of folded algebraic geometry (AG) codes and multiplicity AG codes for constructing subspace-design codes over constant-sized alphabets, alternative pathways must be developed that successfully leverage the structural properties of algebraic function fields. Standard algebraic geometry codes—where rational places are chosen as evaluation points—do not, by themselves, possess the strong ``local properties" required to serve as subspace-design codes. In particular, their zero-coordinate distribution does not satisfy the subspace-design condition, which is essential for achieving optimal list-decoding and proximity-gap guarantees.

To address this limitation, one may naturally modify standard AG codes by taking places of higher degree as evaluation points, rather than restricting to rational places. This modification yields an \emph{additive} algebraic geometry code, i.e., a code that is linear over the base field $\mathbb{F}_q$ even though the alphabet is an extension field $\mathbb{F}_{q^s}$. The use of higher-degree places introduces additional algebraic structure that can be exploited to improve the local properties of the code.

 To analyze the parameters of such a code as a subspace-design code, we construct a Moore matrix associated with the evaluation points. The Moore matrix is a natural tool for studying linear independence over finite fields and plays a crucial role in establishing the subspace-design property. Specifically, given a set of functions, one can form a Moore matrix whose determinant can be shown to be nonzero whenever the corresponding functions are linearly independent over $\mathbb{F}_q$. This determinant argument provides a mechanism for bounding the dimension of subspaces of codewords that vanish on a large fraction of coordinates, which is precisely the condition required for the subspace-design property.

\section{Preliminaries}
\subsection{Function fields}
In this subsection, we briefly introduce some  necessary results from function fields for constructing our multiplicity \AG code.
For convenience of the reader, we start with some background on global function fields over finite fields. The reader may refer to \cite{St09,TV91,NX13} for detailed background on function fields and algebraic-geometric codes.

\subsubsection{General background on function fields}
For a prime power $q$, let $\F_q$ be the finite field of $q$ elements. An { algebraic function field}  over
$\F_q$ in one variable is a field extension $F \supset \F_q$ such
that $F$ is a finite algebraic extension of  $\F_q(x)$ for some
$x\in F$ that is transcendental over $\F_q$. The field $\F_q$ is called the full constant field of $F$ if the algebraic closure of $\F_q$  in $F$ is $\F_q$ itself. Such a function field is also called a global function field. From now on, we always denote by  $F/\F_q$  a function field $F$ with the full constant field $\F_q$.

\subsubsection{Valuations, Places, and Divisors}
A discrete valuation of $F/\F_q$ is a map from $F$ to $\ZZ\cup\{+
\infty\}$ satisfying certain properties (see \cite[Definition 1.1.9]{St09}). Then each discrete valuation $\nu$ from $F/\F_q$ to $\ZZ\cup\{+
\infty\}$ defines a valuation ring $\mO=\{f\in F:\; \nu(f)\ge 0\}$ that is a local ring \cite[Theorem 1.1.13]{St09}. The maximal ideal $P$ of $\mO$ is  given by $P=\{f\in F:\; \nu(f)> 0\}$ and it is called a {\it place}. We denote  the valuation $\nu$ and the local ring $\mO$ corresponding to $P$ by $\nu_P$ and $\mO_P$, respectively. The residue class field $\mO_P/P$, denoted by $F_P$, is a finite extension of $\F_q$. The extension degree $[F_P:\F_q]$ is called the {\it degree} of $P$, denoted by $\deg(P)$. A place of degree one is called a {\it rational} place.

Let $\PP_F$ denote the set of places of $F$. The divisor group, denoted by ${\rm Div}(F)$, is the free abelian group generated by all places in $\PP_F$. An element $G=\sum_{P\in\PP_F}n_PP$ of ${\rm Div}(F)$ is called a divisor of $F$, where $n_P=0$ for almost all $P\in\PP_F$. We denote $n_P$ by $\nu_P(G)$. The degree of $G$ is defined by $\deg(G)=\sum_{P\in\PP_F}n_P\deg(P)$. The support, denoted by $\Supp(G)$, of $G$ is the set $\{P\in\PP_F:\; n_P\neq 0\}$. Thus, $\Supp(G)$ of a divisor $G$ is always a finite subset of $\PP_F$. We say that a divisor $G \ge 0$ if $\nu_P(G) \ge 0$ for all $P \in \PP_F$.

For a nonzero function $z\in F$, the principal divisor  of $z$ is defined to be
${\rm div}(z)=\sum_{P\in\PP_F}\nu_P(z)P$. The principal divisor ${\rm div}(z)$ is indeed a divisor as $z$ has only finitely many zeros and poles.
The zero and pole divisors of $z$ are defined to be ${\rm div}(z)_0=\sum_{\nu_P(z)>0}\nu_P(z)P$ and ${\rm div}(z)_{\infty}=-\sum_{\nu_P(z)<0}\nu_P(z)P$, respectively. Then, we have $\deg({\rm div}(z))=0$, i.e, $\deg({\rm div}(z)_0)=\deg({\rm div}(z)_\infty)$.
For two functions $f,g\in F$ and a place $P$, we have $\nu_P(f+g)\ge \min\{\nu_P(f),\nu_P(g)\}$ and the equality holds if $\nu_p(f)\neq\nu_P(g)$ (note that $\nu_P(0)=+\infty$). This implies that $f+g\neq 0$ if $\nu_P(f)\neq\nu_P(g)$.

The weak approximation theorem gives the following result.
\begin{lemma}\label{lem:2.1} (see \cite[Theorem 1.3.1]{St09}) Let $F/\F_q$ be a function field. Then for any pairwise distinct places $P_1,\dots,P_n$ of $F$, any elements $\Ga_1,\dots,\Ga_n\in\F_q$ and any integers $r_1,\dots,r_n$, there exists a function $z\in F$ such that $\nu_{P_i}(z-\Ga_i)=r_i$ for all $1\le i\le n$. 
\end{lemma}

If $F$ is the rational function field $\F_q(x)$, then every discrete valuation of $F/\F_q$  is given by  either $\nu_{\infty}$ or $\nu_{p(x)}$ for an irreducible polynomial $p(x)$, where $\nu_{\infty}$ is defined by $\nu_{\infty}(f/g)=\deg(g)-\deg(f)$  and $\nu_{p(x)}(f/g)=a-b$ with $p(x)^a||f$ and  $p(x)^b||g$ for two nonzero polynomials $f,g\in\F_q[x]$. It is straightforward to verify that the degrees of places   corresponding to $\nu_{\infty}$ and  $\nu_{p(x)}$ are $1$ and $\deg(p(x))$, respectively.

\subsubsection{Riemann-Roch spaces}\label{subsub:2.1.3} For a divisor $G$ of $F/\F_q$, we define the \emph{Riemann-Roch space} associated with $G$ by
\[\mL(G):=\{f\in F^*:\; {\rm div}(f)+G\ge 0\}\cup\{0\},\]
where $F^*$ denotes the set of nonzero elements of $F$.
Then $\mL(G)$ is a finite dimensional space over $\F_q$ and its
dimension $\ell(G)$ is determined by the Riemann-Roch theorem which
gives
\[\ell(G)=\deg(G)+1-\g+\ell(W-G),\]
where $\g$ is the genus of $F$ and $W$ is a canonical divisor of degree $2\g-2$. Therefore, we
always have that $\ell(G)\ge \deg(G)+1-\g$ and the equality holds
if $\deg(G)\ge 2\g-1$ \cite[Theorems 1.5.15 and 1.5.17]{St09}.

\subsection{Algebraic-geometric codes} Let $\PP=\{P_1,P_2,\dots,P_n\}$ be a set of $n$ distinct places of degree $s\ge 1$ in a function field $F/\F_q$ of genus $\g$. Let $G$ be a divisor of $F$ with $\Supp(G)\cap\PP=\emptyset$. For a function $f\in \mL(G)$, it belongs to the ring $\mO_{P_i}$ for all $1\le i\le n$. Denote by $f(P_i)$ the residue $\bar{f}$ in the residue class field $F_{P_i}=\mO_{P_i}/P_i$.  We define the
algebraic-geometric code 
\begin{equation}
\label{eq:ag-code-defn}
C(\PP,G):=\{(f(P_1),f(P_2),\dots,f(P_n)):\; f\in\mL(G)\}\subseteq F_{P_1}\times\cdots\times F_{P_n}.
\end{equation}
As $F_{P_i}$ is isomorphic to $\F_{q^s}$, we can identify $F_{P_i}$ with $\F_q^s$ with a fixed $\F_q$-isomorphism between $F_{P_i}$ and $\F_q^s$. Thus, $\mC(\PP,G)$ can be viewed as a code in $(\F_q^s)^n$. 

If $s=1$, the code $\mC(\PP,G)$  is the standard algebraic geometry code defined in \cite{St09,TV91}. 

The (generalized) Reed-Solomon codes can be realized under the above framework of algebraic-geometric codes. More precisely speaking, the Reed-Solomon codes are algebraic-geometric codes based on rational function fields. Let us give the detail on the construction of the Reed-Solomon codes
under the framework of algebraic-geometric codes.

Let $F=\F_q(x)$ be a rational function field. Let $\Ga_1,\Ga_2,\dots,\Ga_n$ be $n$ distinct elements of $\F_q$. Denote by $P_i$ the unique zero of $x-\Ga_i$ for $1\le i\le n$ and put $\PP=\{P_1,P_2,\dots,P_n\}$. Let $\Pin$ be the unique pole of $x$. Put $G=(k-1)\Pin$. Then the Riemann-Roch space $\mL(G)$ is the $\F_q$-space consisting of polynomials of degree less than $k$. By definition, we have
\begin{eqnarray*}
\mC(\PP,G)&=&\{(f(P_1),f(P_2),\dots,f(P_n)):\; f\in\mL(G)\}\\
&=&\{(f(\Ga_1),f(\Ga_2),\dots,f(\Ga_n)):\; f\in\F_q[x],\; \deg(f)\le k-1\}.
\end{eqnarray*}

\subsubsection{Local expansions}
\label{sec:2.2.1}
Similar to the Laurent series expansion of a complex function $f(z)$ in the neighborhood of a complex number, one can write functions in a function field as a power series (with finitely many negative powers) around a place $P$, called the \emph{local expansion around $P$}.

Let $F/\F_q$ be a function field and let
$P$ be a rational place. An element $t$ of $F$ is called a local parameter at $P$ if $\nu_P(t)=1$ (such a local parameter always exists) --- intuitively this is a function which has a simple zero at $P$, similar to how $(z-1)$ has a simple zero at $1$.
For a nonzero function $f\in F$ with $\nu_P(f)\ge
v$, we have
$\nu_P\left(\frac f{t^v}\right)\ge 0.$
Put
$f_v=\left(\frac f{t^v}\right)(P),$
i.e., $f_v$ is the value of the function $f/t^v$ at $P$.  Note that the function $f/t^v-f_v$ satisfies
$\nu_P\left(\frac f{t^v}-f_v\right)\ge 1,$
hence we know that
$ \nu_P\left(\frac {f-f_vt^v}{t^{v+1}}\right)=\nu_P\left(\frac {f-f_vt^v}{t^{v}}\right)-1\ge 0.$
Put
$f_{v+1}=\left(\frac{f-f_vt^v}{t^{v+1}}\right)(P).$ Note that the function $\frac{f-f_vt^v}{t^{v+1}}-f_{v+1}$ satisfies $\nu_P\left(\frac{f-f_vt^v}{t^{v+1}}-f_{v+1}\right)\geq1.$
Then  $\nu_P(f-f_vt^v-f_{v+1}t^{v+1})\ge v+2$.

Assume that we have obtained a sequence $\{f_r\}_{r=v}^m$ ($m>v$)
of elements of $\F_q$ such that
$\nu_P(f-\sum_{r=v}^kf_rt^r)\ge k+1$
for all $v\le k\le m$.
Put
$f_{m+1}=\left(\frac{f-\sum_{r=v}^mf_rt^r}{t^{m+1}}\right)(P).$
Then  $\nu_P(f-\sum_{r=v}^{m+1}f_rt^r)\ge m+2$.
In this way we continue our construction of  $f_r$. Then we obtain
an infinite sequence $\{f_r\}_{r=v}^{\infty}$ of elements of $\F_q$ such
that
$
\nu_P(f-\sum_{r=v}^mf_rt^r)\ge m+1
$
for all $m\ge v$.
We summarize the above  construction in the formal expansion
\begin{equation}
\label{eq:x1}
f=\sum_{r=v}^{\infty}f_{r}t^r,
\end{equation}
which is called the \emph{local expansion} of $f$ at $P$.

It is clear that the local expansion of a function depends on the choice of the local
parameter $t$. Note that if a power series
$\sum_{i=v}^{\infty}f_it^i$ satisfies $\nu_P(f-\sum_{i=v}^mf_it^i)\ge
m+1$ for all $m\ge v$, then it is a local expansion of $f$.

Let $f$ belong to a Riemann-Roch space $\mL(G)$ with $\deg(G)=d$. Denote $\nu_P(G)$ by $v$,  then the first $d+1$ coefficients $f_{v},f_{v+1},\dots,f_{v+d}$ in \eqref{eq:x1} determine the function $f$. To see this, assume that $g$ is a function of $\mL(G)$ with the  first $d+1$ coefficients in its local expansion equal to those of $f$. Then we have $f-g\in\mL(G-(d+1)P)$ which is the zero vector space, since $\deg(G-(d+1)P)=d-(d+1)=-1<0$, we have $\mL(G-(d+1)P)=\{0\}$. This implies that $f=g$.

\subsubsection{Differentials}
An element $z$ of a function field $F/\F_q$ is called {\it separating} if the extension $F/\F_q(z)$ is a separable extension. Note that a local parameter at a place is a separating element (see \cite[Subsection 4.1]{St09}). It is well known that  $z$ is a separating element of $F$ if and only if the differential $dz$ is not zero (see \cite[Lemma 4.1.6]{St09}). 
The differentials of $F$ from a vector space $\GO_F$ over $F$ of dimension $1$. In other words, $\GO_F=F\cdot dz$ for  any separating element $z$ of $F$. 

We now define derivatives of functions of $F$. Fix a separating element $z$ of $F$. For a function $f$ of $F$, we define the derivative of $f$ with respect to $z$ to be the unique function $g$ satisfying $df=gdz$. We denote $g$ by $\Gd_z(f)$, or simply by $f^{(1)}$ if there is no confusion. Denote $f$ by $f^{(0)}$.  Inductively, for $m\ge 1$, we define the $m$th derivative of $f$ by $\Gd_z(f^{(m-1)})$, denoted by $f^{(m)}$.  Then we have the following result.
\begin{lemma}\label{lem:2.2}
    One has  
    \begin{itemize}
       \item[{\rm (i)}] $(fg)^{(1)}=fg^{(1)}+gf^{(1)}$.
    \item[{\rm (ii)}] 
    $(\Ga f+\Gb g)^{(m)}= \Ga f^{(m)}+ \Gb g^{(m)}$ for any $m\ge 0$, any functions $f,g\in F$ and $\Ga,\Gb\in\F_q$.
     \end{itemize}
\end{lemma}
\begin{proof}
(i) follows from \cite[Proposition 4.1.8(b)]{St09}.

For (ii), we first show that $(\Ga f+\Gb g)^{(1)}= \Ga f^{(1)}+ \Gb g^{(1)}$. By definition, we have $(\Ga f+\Gb g)^{(1)}dz=d(\Ga f+\Gb g)=\Ga df+\Gb dg=\Ga f^{(1)}dz+\Gb g^{(1)}dz$. This gives the desired result.

    We prove it by induction on $m$. It is clearly true for $m=0$. 
Suppose that it is true for $m-1$. Then we have $(\Ga f+\Gb g)^{(m)}dz=d(\Ga f+\Gb g)^{(m-1)}=d(\Ga f^{(m-1)}+\Gb g^{(m-1)})=\Ga df^{(m-1)}+\Gb dg^{(m-1)}=\Ga f^{(m)}dz+\Gb g^{(m)}dz$. This completes the induction. 
\end{proof}

Let $t$ be a local parameter at a place $P$ of $F$, define the valuation of a differential $\omega \in\Omega_F$ at $P$, denoted $\nu_P(\omega)$, to be $\nu_P(f)$, where $f$ is the unique function satisfying $\Go=fdt$. The local expansion of $\Go=fdt$ is defined to be the $(\sum_{i=r}^{\infty}f_it^i)dt$ if $f$ has the local expansion $f=\sum_{i=r}^{\infty}f_it^i$. We have the following result. In particular, if $\nu_P(f)\ge 0$, then $\nu_P(df)\ge 0$. 
\begin{lemma}[see Proposition 4.2.7 of \cite{St09}]\label{lem:2.3}  Let $P$ be a rational place of $F$ and let $f\in F$. If $f$ has local expansion $f=\sum_{i=r}^{\infty}f_it^i$ at $P$ for some $r\in\ZZ$ and $f_i\in \F_q$, then $df$ has the local expansion $df=\sum_{i=r}^{\infty}if_it^{i-1}dt$ at $P$. Consequently, the derivative $f^{(1)}=\Gd_t(f)$ with respective to $t$ is $f^{(1)}=\sum_{i=r}^{\infty}if_it^{i-1}$.
\end{lemma}

\begin{ex}{\rm  Consider the rational function field $F=\F_q(x)$. Then $\Gd_x(f)$ is the usual derivative. If $z=x^{q-1}-1$, then we have $\Gd_z(f)=\frac{df}{dz}=\frac{df}{(q-1)x^{q-2}dx}$, i.e., $\Gd_x(f)=\frac{df}{dx}=(q-1)x^{q-2}\Gd_z(f)=-x^{q-2}\Gd_z(f)$ since $q-1\equiv -1\in\F_q$.

}    
\end{ex}
\subsection{Garcia-Stichtenoth tower}\label{subsec:2.3}

\label{subsec:gs-prelim}
Let $q$ be a prime power and let $s\ge 2$ be an even integer. Put $r=q^{s/2}$. The Garcia-Stichtenoth towers that
we are going to use for our code construction were discussed in
\cite{GS95}. The reader may refer to \cite{GS95} for the
detailed background on the Garcia-Stichtenoth function tower.  There
 are two optimal Garcia-Stichtenoth towers that are equivalent. For
simplicity, we introduce the tower defined by the following recursive
equations \cite{GS95} over $\F_{q}$
\begin{equation}\label{eq:3}
x_{i+1}^r+x_{i+1}=\frac{x_i^{r}}{x_i^{r-1}+1},\quad i=1,2,\dots,e-1.
\end{equation}
Put  $K_e=\F_q(x_1,x_2,\dots,x_{e})$ for $e\ge 2$.

\subsubsection{Genus}
The genus $\g_e$ of the function field $K_e$ is given by
\[\g_e=\left\{\begin{array}{ll}
(r^{e/2}-1)^2&\mbox{if $e$ is even}\\
(r^{(e-1)/2}-1)(r^{(e+1)/2}-1)&\mbox{if $e$ is odd.}\end{array}
\right.\]
Thus the genus $\g_e$ is at most $r^e$. (Compare this with the genus of the Hermitian tower, which grows like $er^e$ where $e \leqslant\frac{r}{2} = \frac{q^{s/2}}{2}$; this smaller genus is what allows us to pick $e$ as large as we want in the Garcia-Stichtenoth tower, while keeping the field size $q$ fixed.)

\subsubsection{Places of degree $s$}
The function field $K_e$ is defined over $\F_q$. To compute the number of places of degree $s$ in $K_e$, we consider the constant extension $F_e:=\F_{r^2}\cdot K_e$. $F_e$ has at least $r^{e-1}(r^2-r)+1$ $\F_{q^s}$-rational places. One of these is the ``point at infinity" which is the unique pole $\Pin$ of $x_1$ (and is fully ramified). The other $r^{e-1}(r^2-r)$ come from the $\F_{q^s}$-rational places lying over the unique zero of $x_1-\Ga$ for each $\Ga\in\F_{q^s}$ with $\Ga^r+\Ga\not=0$. Note that for every $\Ga\in\F_{q^s}$ with $\Ga^r+\Ga\not=0$, the unique zero of $x_1-\Ga$ splits completely in $K_e$, i.e., there are $r^{e-1}$ $\F_{q^s}$-rational places lying over the zero of $x_1-\Ga$. Let $\PP$ be the set of all the $\F_{q^s}$-rational places lying over the zero of $x_1-\Ga$ for all $\Ga\in\F_{q^s}$ with $\Ga^r+\Ga\not=0$.
Then, intuitively, one can think of the $r^{e-1}(r^2-r)$ $\F_{q^s}$-rational places in $\PP$ as being given by $e$-tuples $(\Ga_1,\Ga_2,\dots,\Ga_e)\in \F_q^e$ that satisfy $\Ga_{i+1}^r+\Ga_{i+1}=\frac{\Ga_i^{r}}{\Ga_i^{r-1}+1}$ for $i=1,2,\dots,e-1$ and $\Ga_1^r+\Ga_1\neq 0$. For each value of $\Ga\in\F_{q^s}$, there are precisely $r$ solutions to $\beta \in\F_{q^s}$ satisfying $\beta^r + \beta = \frac{\Ga^{r}}{\Ga^{r-1}+1}$, so the number of such $e$-tuples is $r^{e-1}(r^2-r)$ ($r^2-r$ choices for $\Ga_1$, and then $r$ choices for each successive $\Ga_i$, $2 \le i  \le e$). In conclusion, the number $B_s$ of places of  degree $s$ of $K_e$ satisfies $\frac{sB_s}{\g_e}\rightarrow r-1=q^{s/2}-1$.

\subsubsection{Riemann-Roch spaces}
As shown in \cite{Shum01}, every function of $K_e$ with a pole only at $\Pin$ has an expression of the form
\begin{equation}\label{eq:x6}x_1^a\left(\sum_{i_1=0}^{(e-2)r+1}
\sum_{i_2=0}^{r-1}\cdots\sum_{i_e=0}^{r-1}c_{\bi}h_1\frac{x_1^{i_1}x_2^{i_2}\cdots x_e^{i_e}}{\pi_2\dots\pi_{e-1}}\right),
\end{equation}
where $a\ge 0, c_{\bi}\in\F_q$, and for $1\le j<e$ (${\bf i}=(i_1,\dots,i_e)$), $h_j=x_j^{r-1}+1$ and $\pi_j=h_1h_2\dots h_j$. Moreover, Shum et al. \cite{Shum01} present an algorithm running in time polynomial in $l$  that outputs a basis  over $\F_q$ of $\cL(l\Pin)$   in the above form.

\subsection{Subspace design codes}\label{subsec:2.4}
Subspace-design codes were introduced in \cite{GX13} as a way to pre-code certain algebraic codes to enable list decoding with small lists. The core idea is to ensure that for any low-dimensional subspace of codewords, the average dimension of its intersection with the coordinate-wise zero subspaces is tightly controlled. This property, known as the \emph{subspace-design property}, has proven to be remarkably powerful.

Recent work \cite{GG25} has established that subspace-design codes with near-optimal parameters exhibit \emph{optimal proximity gaps}—a property crucial for the soundness of interactive oracle proofs (IOPs) used in SNARKs. Specifically, they show that such codes have proximity gaps for proximity parameters approaching the optimal $1 - R$  bound, where $ R$   is the code rate.

Let \(\mathbb{F}_q\) be a finite field. A code \(\mathcal{C} \subseteq (\mathbb{F}_q^s)^n\) is called an \(\mathbb{F}_q\)-additive code if it is $\F_q$-linear. 

\begin{defn}[Subspace-Design Property \cite{GG25}]{\rm Let $\tau$ be a function from $\mathbb{N}$ to $[0,1]$ and let $r\in\mathbb{N}$. 
An \(\mathbb{F}_q\)-additive code \(\mathcal{C} \subseteq (\mathbb{F}_q^s)^n\) is said to be a $(\tau(r),r)$-subspace design code if for every \(\mathbb{F}_q\)-linear subspace \(\mathcal{A} \subseteq \mathcal{C}\) of dimension at most \(r\),
\[
\frac{1}{n}\sum_{i=1}^n \dim (\mathcal{C}_i\cap \mA) \le \dim(\mA) \cdot \tau,
\]
where $\mathcal{C}_i=\{(\bc_1,\dots,\bc_n)\in\mC:\; \bc_i=\bo\}$
}\end{defn}

Intuitively, the subspace-design property says that for any low-dimensional subspace \(\mathcal{A}\), the average dimension of \(\mathcal{C}_i\cap \mA\) across coordinates is bounded by \(\dim(\mathcal{A}) \cdot \tau\). For a rate \(R\) code, we necessarily have \(\tau \ge R - 1/n\) for all \(r\).

\subsection{List recovery and proximity gap}\label{subsec:2.5}

\paragraph{List recovery.}
It was shown in \cite{BCDZ25a,GG25} that subspace design codes have list-recovery capacity and proximity gap and correlated agreement.

\begin{lemma}[Subspace design $\Rightarrow$ list recovery
\cite{BCDZ25a,GG25}]\label{lem:2.5}
Let $\mC\subseteq(\F_q^s)^n$ be a $(\tau,r)$-subspace-design additive code.
For every input-list size $\ell\in\mathbb{N}$ and every $\varepsilon>0$,  $\mC$ is
\[
 \left(1-\tau\!\left(\left\lceil\frac{\ell}{\varepsilon}\right\rceil\right)
       -\varepsilon,\ \ell,\ L\right)
 \text{-list recoverable},
\]
where
\[
 L\leq
 \left(
   \frac{\ell}{
   \tau\!\left(\left\lceil\ell/\varepsilon\right\rceil\right)+\varepsilon}
 \right)^{
 \left(
   \tau\!\left(\left\lceil\ell/\varepsilon\right\rceil\right)+\varepsilon
 \right)/\varepsilon}.
\]
Equivalently, for all $S_1,\ldots,S_n\subseteq\Sigma$ with
$|S_i|\leq\ell$,
\[
 \left|\left\{\bc=(c_1,\dots,c_n)\in\mC:
 \bigl|\{i:c_i\in S_i\}\bigr|>
 \left[\tau\!\left(\left\lceil\frac{\ell}{\varepsilon}\right\rceil\right)
       +\varepsilon\right]n
 \right\}\right|\leq L.
\]
\end{lemma}

The reduction therefore maps the parameters as follows:
\begin{equation}\label{eq:map}
 \boxed{
 r=\left\lceil\frac{\ell}{\varepsilon}\right\rceil,
 \qquad
 \rho=1-\tau(r)-\varepsilon,
 \qquad
 L\leq\left(\frac{\ell}{\tau(r)+\varepsilon}\right)^{
                    (\tau(r)+\varepsilon)/\varepsilon}.}
\end{equation}

We now give a definition for proximity gap.
\begin{defn}[Proximity Gap with slack]
{\rm An \(\mathbb{F}_q\)-additive code \(\mathcal{C} \subseteq (\F_q^s)^n\) is said to have \((\delta,\Ge,\gamma)\) proximity gap for lines if for all \(\bu_0,\bu_1\in (\F_q^s)^n\), the following implication holds:
\[
\Pr_{\alpha \in \F_q}\left[ \Delta\!\left(\bu_0+\Ga\bu_1,\; \mathcal{C}\right) \le \delta \right] > \Ge
\quad\Longrightarrow\quad
\forall \alpha \in \mathbb{F},\;
\Delta\!\left(\bu_0+\Ga\bu_1,\; \mathcal{C}\right) \le \frac{\delta}{1-\gamma}.
\]}
\end{defn}

We now introduce a stronger notion--correlated Agreement.
\begin{defn}[Correlated Agreement]
{\rm An \(\F_q\)-additive code \(\mathcal{C} \subseteq \Sigma^n\) is said to have \emph{\((\delta,\Ge,\gamma)\) correlated agreement} if for all \(\bu_0,\bu_1\in \Sigma^n\), whenever
\[
\Pr_{\alpha \in \F_q}\left[ \Delta\!\left(\bu_0+\Ga\bu_1,\; \mathcal{C}\right) \le \delta \right] > \Ge,
\]
there exist codewords \(\bc_0,\bc_1\in \mathcal{C}\) such that
\[
\left|\left\{ i \in [n] \;:\; \left( c_{0,i} \neq u_{0,i} \right)\bigvee\left( c_{1,i} \neq u_{1,i}\right)  \right\}\right| \le \frac{\delta n}{1-\gamma}.
\]
It is clear that  \(\mathcal{C}\) has \((\delta,\Ge,\gamma)\) proximity gap as long as it has  \((\delta,\Ge,\gamma)\) correlated agreement.

}\end{defn}

\begin{lemma}[Correlated Agreement for Subspace-Design Codes \cite{GG25}]\label{lem:2.6}
For any integers \(r, m \in \mathbb{N}\) and any \(\tau\)-subspace-design code \(\mathcal{C} \subseteq (\F_q^s)^n\), the code has correlated agreement
\[
\left( 1 - \tau(r) - \frac{3}{r}, \frac{r (m(1 - \tau(r)) + r^2)}{2q}, \frac{1}{m} \right).
\]
\end{lemma}

\subsection{Moore matrix}
  \label{subsec:2.6}
  Let $F$ be an extension of $\F_q$. For elements $\Ga_1,\dots,\Ga_r\in F$, the Moore matrix is defined by
  \[M(\Ga_1,\dots,\Ga_r)=\left(\begin{array}{cccc}
\Ga_1&\Ga_2&\cdots&\Ga_r\\
\Ga_1^q&\Ga_2^q&\cdots&\Ga_r^q\\
\vdots&\vdots&\ddots&\vdots \\
\Ga_1^{q^{r-1}} &\Ga_2^{q^{r-1}} &\cdots&\Ga_r^{q^{r-1}}
\end{array}
\right)\in F^{r\times r}.\]
  The determinant $\det(M(\Ga_1,\dots,\Ga_r))$ is given by the following formula
  \[\det(M(\Ga_1,\dots,\Ga_r))=\prod_{(c_1,\dots,c_r)}(c_1\Ga_1+\cdots+c_r\Ga_r),\]
  where $(c_1,\dots,c_r)$ runs through all non-zero direction vectors in $\F_q^r$. Thus, $\det(M(\Ga_1,\dots,\Ga_r))\neq 0$ if and  only if  $\Ga_1,\dots,\Ga_r$ are $\F_q$-linearly independent.
\section{Subspace design codes over constant alphabets}

Let $F/\F_q$ be a function field of genus $\g$. Assume that $\Pin$ is a rational place of $F$ and choose   $n$ pairwise distinct places $P_1,\dots,P_n$ of degree $s\ge 1$. Put $\PP=\{P_1,P_2,\dots,P_n\}$ and define the divisor $G=(k+\g-1)\Pin$. 
\begin{lemma}\label{lem:3.1} Let $\PP$ and divisor $G$ defined above. If $ns>k+\g-1$, then the code $\mC(\PP,G)\subseteq(\F_q^s)^n$ is additive over $\F_q$ with rate $R\ge\frac{k}{ns}$ and the minimum distance at least $n-\frac {k+\g-1}s$. 
\end{lemma}
\begin{proof}
 For any $\Ga,\Gb\in\F_q$ and $f,g\in\mL((k+\g-1)\Pin)$, we have $(\Ga f+\Gb g)(P_i)=\overline{\Ga f+\Gb g}=\Ga \bar{f}+\Gb \bar{g}=\Ga f(P_i)+\Gb g(P_i)$. Hence,   $\mC(\PP,G)$ is additive.

 If $f\in\mL((k+\g-1)\Pin)$ is a nonzero function, and let $I=\{i\in[n]: f(P_i)\neq 0\},\quad [n]\setminus I=\{i: f(P_i)=0\}\subset[n]$ be the support set of $(f(P_1),\dots,f(P_n))$, then $\nu_{P_i}(f)\ge 1$ for all $i\in [n]\setminus I$. This implies that $f$ is a nonzero function of the Riemann-Roch space $\mL((k+\g-1)\Pin-\sum_{i\in [n]\setminus I}P_i)$. Hence, $\deg((k+\g-1)\Pin-\sum_{i\in [n]\setminus I}P_i)\ge 0$, i.e., $k+\g-1\ge (n-|I|)s$. This gives $|I|\ge n-\frac{k+\g-1}s$. This means that  the minimum distance of $\mC(\PP,G)$  at least $n-\frac {k+\g-1}s$. As $n-\frac {k+\g-1}s>0$, the $\F_q$-dimension of $\mC(\PP,G)$ is equal to the $\F_q$-dimension of $\mL((k+\g-1)\Pin)$ which is at least $k$ by the Riemann-Roch Theorem.  
\end{proof}
We now show that $\mC(\PP,G)$ given in this section is a subspace design code.
\begin{theorem}\label{thm:3.2} Let $F/\F_q$ be a function field of genus $\g$. Assume that $\Pin$ is a rational place of $F$ and choose   $n$ pairwise distinct places $P_1,\dots,P_n$ of degree $s\ge 1$. Put $\PP=\{P_1,P_2,\dots,P_n\}$ and define the divisor $G=(k+\g-1)\Pin$. If $n>\frac{k+\g-1}s$,
then    $\mC(\PP,G)$ is a $(\tau,r)$-subspace design code with
    \[\tau=1+\frac1r\log_q \left[\left(1-\frac1{q^r}\right)\left(R+\frac{\g-1}{ns}\right)+\frac1{q^r}\right].\]
\end{theorem}
\begin{proof}
Let $\pi$  be the map from $\mL((k+\g-1)\Pin)$  to $\mC(\PP,G)$ given by $f\mapsto (f(P_1),\dots,f(P_n))$. By Lemma \ref{lem:3.1}, $\pi$ is an $\F_q$-isomorphism. Let $\mA$ be an $\F_q$-subspace of  $\mC(\PP,G)$ and let $\mC_i$ be the subspace of $\mC(\PP,G)$ with the $i$th position equal to $\bo$. Let $W=\pi^{-1}(\mA)$ and $W_i=\pi^{-1}(\mA\cap \mC_i)\subseteq W$. Then $\dim_{\F_q}(W)=\dim_{\F_q}(\mA)$ and $\dim_{\F_q}(W_i)=\dim_{\F_q}(\mA\cap \mC_i)$. 

Put $r=\dim_{\F_q}(W)$ and $r_i=\dim_{\F_q}(W_i)$. 
Fix an $\F_q$-basis $\{f_1,\dots,f_r\}$ of $W$. Let $\{g_1,\dots,g_{r_i}\}$ be an $\F_q$-basis of $W_i$. Extend this basis to an $\F_q$-basis $\{g_1,\dots,g_{r}\}$ of $W$. Then there exists an invertible matrix $A_i\in\F_q^{r\times r}$ such that
\[(f_1,\dots,f_r)=(g_1,\dots,g_r)A_i.\]
As every element of $A_i$ is an element of $\F_q$, raising each $f_i$ to $q^j$th power gives 
\[(f_1^{q^j},\dots,f_r^{q^j})=(g_1^{q^j},\dots,g_r^{q^j})A_i.\]
Combining the above equations for $j=0,1,\dots, r-1$ gives the following matrix equation 
\begin{equation}\label{eq:}
   \begin{pmatrix}
    f_1&f_2&\cdots&f_r\\
f_1^{q}&f_2^{q}&\cdots&f_r^{q}\\
\cdots&\cdots&\cdots&\cdots
\\
    f_1^{q^{r-1}}&f_2^{q^{r-1}}&\cdots&f_r^{q^{r-1}}
\end{pmatrix}=\begin{pmatrix}
    g_1&g_2&\cdots&g_r\\
g_1^{q}&g_2^{q}&\cdots&g_r^{q}\\
\cdots&\cdots&\cdots&\cdots
\\
    g_1^{q^{r-1}}&g_2^{q^{r-1}}&\cdots&g_r^{q^{r-1}}
\end{pmatrix}A_i.
\end{equation}
As $f_1,\dots,f_r$ are $\F_q$-linearly independent, the determinant $M(f_1,\dots,f_r)\neq 0$. If is clear that the determinate $M(f_1,\dots,f_r)$ belongs to the Riemann-Roch space \[\mL\left((\sum_{j=0}^{r-1}q^j)(k+\g-1)\Pin\right)=\mL\left(\frac{q^r-1}{q-1}(k+\g-1)\Pin\right).\] 
Note that every term of the determinant $M(g_1,\dots,g_r)$ contains factor $\prod_{j=1}^{r_i}g_j^{q^{\ell_j}}$ for some $0\le \ell_1,\dots,\ell_{r_i}\le r-1$. Moreover $\ell_1,\dots,\ell_{r_i}$ are pairwise distinct. This implies that $\sum_{j=1}^{r_i}q^{\ell_j}\ge \sum_{j=0}^{r_i-1}{q^j}$. 

Since $P_i$ is a zero of $g_j$ for $1\le j\le r_1$,  $P_i$ is a zero of $g_j^{q^{\ell_j}}$ with multiplicity $q^{\ell_j}$. Therefore, $P_i$ is a zero of $\prod_{j=1}^{r_i}g_j^{q^{\ell_j}}$  with multiplicity $\sum_{j=1}^{r_i}q^{\ell_j}\ge \sum_{j=0}^{r_i-1}{q^j}=\frac{q^{r_i}-1}{q-1}$.
This means that  $P_i$ is a zero of $\det(M(g_1,\dots,g_r))$  with multiplicity at least $\frac{q^{r_i}-1}{q-1}$.

On the other hand, as $\det(M(f_1,\dots,f_r))$ is a product of $\det(M(g_1,\dots,g_r))$ with the nonzero constant $\det(A_i)\in\F_q$, $P_i$ is also a zero of $\det(M(f_1,\dots,f_r))$ with multiplicity at least $\frac{q^{r_i}-1}{q-1}$ for $1\le i\le n$. In conclusion, we have 
\[\det(M(f_1,\dots,f_r))\in\mL\left(\frac{q^r-1}{q-1}(k+\g-1)\Pin-\sum_{i=1}^n\frac{q^{r_i}-1}{q-1}P_i\right).\]
As $\det(M(f_1,\dots,f_r))$ is not a zero function, we have
\[\deg\left(\frac{q^r-1}{q-1}(k+\g-1)\Pin-\sum_{i=1}^n\frac{q^{r_i}-1}{q-1}P_i\right)\ge 0,\]
i.e., 
\begin{equation}\label{eq:6}\frac{q^r-1}{q-1}(k+\g-1)\ge s\sum_{i=1}^n\frac{q^{r_i}-1}{q-1}\ge 0.\end{equation} 
Equation \eqref{eq:6} gives 
\[
 (q^r-1)(k+\g-1)\ge s\sum_{i=1}^nq^{r_i}-ns
 \ge sn\left(\prod_{i=1}^n q^{r_i}\right)^{1/n}-ns.\]
 Hence,
 \[ (q^r-1)\left(R+\frac{\g-1}{ns}\right)+1\ge 
\left(\prod_{i=1}^n q^{r_i}\right)^{1/n}\]
 Taking logarithm of base $q$ for the both sides of the above inequality gives
 \[ r+\log_q \left[\left(1-\frac1{q^r}\right)\left(R+\frac{\g-1}{ns}\right)+\frac1{q^r}\right]\ge\frac1n\sum_{i=1}^nr_i.\]
 The desired result follows.
\end{proof}
Applying the Garcia-Stichtenoth tower given in Subsection 2.3, we obtain the following result.
\begin{cor}
Let $q$ be a prime power, then for any $r\ge1$ and even integer $s\ge2$, there is an $\F_q$-linear $(\tau,r)$-subspace design code of rate $R$ over $(\F_q)^s$ with length tending $\infty$ and 
\[\tau=1+\frac1r\log_q \left[\left(1-\frac1{q^r}\right)\left(R+\frac{1}{q^{s/2}-1}\right)+\frac1{q^r}\right]=\frac1r\log_q \left[1+(q^r-1)\left(R+\frac{1}{q^{s/2}-1}\right)\right].\]
\end{cor}

\begin{rem}{\rm \begin{itemize}
    \item[(i)] By \cite{Shum01}, the algebraic geometry codes given in Theorem \ref{thm:3.2} can be constructed with $O(n^2)$ field operations over $\F_{q^s}$. Thus, encoding of the code requires $\Tilde{O}(sn^2)$ operations.
    \item[(ii)] To obtain a meaningful value of $\tau$, we may take $s$ to be a product of $r$ with a constant. For instance, we can take $s=40r$. 
 Denote by $\Ge(q,r,R)$ the difference $\frac1r\log_q \left[1+(q^r-1)\left(R+\frac{1}{q^{20r}-1}\right)\right]-R$. Then  algebraic geometry codes given in Theorem \ref{thm:3.2} is a $(\tau=R+\Ge(r,R),r)$ subspace design code over
alphabet size $\exp(40r)$. 
\end{itemize}
}\end{rem}

We now get a list decodable codes from our subspace design codes.

\begin{cor}\label{cor:3.3}
For an integer $L\ge 1$ and a real $R\in(0,1)$, define $$\Ge(L,R):=\frac1L\log_q \left[1+(q^L-1)\left(R+\frac{1}{q^{20L}-1}\right)\right]-R.$$ Then the algebraic geometry code given in Theorem \ref{thm:3.2} is $(\frac{L}{L+1}(1-R-\Ge(L,R)),L)$-list decodable over alphabet size $\exp(40L)$. 
\end{cor}
\begin{proof}
    Put  $L=r$ and $s=40L$. The desired result follows from Theorem \ref{thm:3.2} and Lemma \ref{lem:2.5}.
\end{proof}

Combining Lemma \ref{lem:2.6} with Corollary \ref{cor:3.3} gives the following result on proximity gap.
\begin{cor}\label{cor:3.4}
For any integers \(r, m \in \mathbb{N}\) and real $R\in(0,1)$, the algebraic geometry code given in Theorem \ref{thm:3.2} is additive code over $\F_q^{40r}$ and has rate $R$ and proximity gap
\[
\left( 1 -R- \Ge(q,r,R) - \frac{3}{r}, \frac{r (m(1 - R- \Ge(q,r,R)) + r^2)}{2q}, \frac{1}{m} \right).
\]
\end{cor}
\begin{proof}
    Put $L=r$ and $s=40L$. The desired result follows from Corollary \ref{cor:3.3} and Lemma \ref{lem:2.5}.
\end{proof}
\section{Multiplicity \AG codes}
\subsection{Wronskian}
Before presenting our construction of subspace design codes, we need to generalize Wronskian from polynomials to functions. 

\begin{lemma}\label{lem:4.1} Let $P$ be a rational place of $F$ with a local parameter $t-\Ga$ for some function $t\in F$ and $\Ga\in\F_q$. Let $f$ be a function of $F$ with $\nu_P(f)\ge0$.  If $r$ is a positive integer less than or equal to the characteristic $p$ of $F$, then $\nu_P(f)\ge r$ if and only if $f^{(i)}(P)=0$ for all $i=0,1,\dots,r-1$, where $f^{(i)}$ is the $i$th derivative with respect to $t$.
\end{lemma}
\begin{proof} First of all, by Lemma \ref{lem:2.3}, $f^{(i)}$ is well defined for all $i\ge 0$. Assume that $f$ has a local expansion $f=\sum_{j=0}^{\infty}a_j(t-\Ga)^j$ for some $a_j\in\F_q$ and a local parameter $t-\Ga$ at $P$.  It follows from Lemma \ref{lem:2.3} that $f^{(i)}$ has a local expansion $f^{(i)}=\sum_{j=i}^{\infty}j(j-1)\cdots(j-i+1)a_j(t-\Ga)^{j-i}$. This gives that $f^{(i)}(P)=i!a_i$. Since $i\le r-1<p$, we have that $i!a_i=0$ if and only if $a_i=0$. This implies that $f^{(i)}(P)=0$ for all $i=0,1,\dots,r-1$  if and only if $a_i=0$ for all $i=0,1,\dots,r-1$. This is equivalent to the fact that $\nu_P(f)\ge r$.
\end{proof}

Let $\Pin,P_1,\dots,P_n$ be $n+1$ pairwise distinct rational places  and let $\Ga_1,\dots,\Ga_n$ be elements of $\F_q$. By the Weak Approximation Theorem \ref{lem:2.1}, there exists an element $t\in F$ such that $\nu_{\Pin}(t)=-1$ and $\nu_{P_i}(t-\Ga_i)=1$ for all $1\le i\le n$.

Let us consider a subspace of the Riemann-Roch space $\mL((2\g+k-1)\Pin)$ for some $k\le p$, where $p$ is the characteristic of $\F_q$.  By the Riemann-Roch Theorem, we have $\ell((2\g+i)\Pin)=2\g+i+1-\g=\g+1+i$ for all $i\ge -1$. This implies that one can find  functions $h_i$ in $\mL((2\g+i)\Pin)\setminus\mL((2\g+i-1)\Pin)$, i.e., $\nu_{\Pin}(h_i)=-2\g-i$  for all $i\ge 0$. We denote by $\mL_k$ the subspace of $\mL((2\g+k-1)\Pin)$ spanned by $\{h_0,h_1,\dots,h_{k-1}\}$:
\begin{equation}\label{eq:8x}
    \mL_k={\rm Span}_{\F_q}\{h_0,h_1,\dots,h_{k-1}\}.
\end{equation}

With the above preparation, we can generalize Wronskian from polynomials to functions. For functions $f_1,\dots,f_r\in F$, define the Wronskian of these functions to be the matrix
\[W(f_1,\dots,f_r)=\begin{pmatrix}
    f_1&f_2&\cdots&f_r\\
f_1^{(1)}&f_2^{(1)}&\cdots&f_r^{(1)}\\
\cdots&\cdots&\cdots&\cdots
\\
    f_1^{(r-1)}&f_2^{(r-1)}&\cdots&f_r^{(r-1)}
\end{pmatrix},\]
where $f^{(i)}$ is the $i$th derivative with respect to $t$ chosen above.

\begin{lemma}\label{lem:4.2} Let $p$ be the characteristic of $\F_q$. Let
 $r,k$ be two integers satisfying $r\le k\le p$. Then the functions   $f_1,\dots,f_r\in \mL_k$ are $\F_q$-linearly independent if and only if the determinant of the Wronskian $W(f_1,\dots,f_r)$ is not zero. 
\end{lemma}
\begin{proof}
 Assume that  the determinant of the Wronskian $W(f_1,\dots,f_r)$ is not zero. Suppose that   $f_1,\dots,f_r\in \mL_k$ were $\F_q$-linearly dependent. Then there exist $\Gl_1,\dots,\Gl_r\in\F_q$ that are not all zero  such that $\sum_{i=1}^r\Gl_if_i=0$. By Lemma \ref{lem:2.2}(ii), we have $\sum_{i=1}^r\Gl_if_i^{(\ell)}=0$ for all $\ell\ge 0$. This implies that the columns of $W(f_1,\dots,f_r)$ are $\F_q$-linearly dependent. Hence, $W(f_1,\dots,f_r)=0$, a contradiction. 

 Now assume that $f_1,\dots,f_r\in \mL_k$ are $\F_q$-linearly independent. Then one can find a new basis $\{h_1,\dots,h_r\}$ of the linear span ${\rm Span}_{\F_q}\{f_1,\dots,f_r\}$ such that $\nu_{\Pin}(h_1)>\nu_{\Pin}(h_2)>\cdots>\nu_{\Pin}(h_r)$.  It is clear that  $W(f_1,\dots,f_r)= W(h_1,\dots,h_r)A$ for an invertible matrix $A\in\F_q^{r\times r}$. Thus, it is sufficient to show that $\det(W(h_1,\dots,h_r))$ is not zero. Assume that $h_i$ has the local expansion $h_i=\sum_{j=-m_i}^{\infty}a_{ij}(1/t)^j=\sum_{j=-\infty}^{m_i}a_{i,-j}t^j$ at $\Pin$, where  $\nu_{\Pin}(h_i)=-m_i$ and $a_{ij}\in\F_q$ with $a_{i,-m_i}\neq0$. Then the term with the highest degree of the determinant of $W(h_1,\dots,h_r)$ is equal to 
 \begin{equation}\label{eq:8} \prod_{i=1}^ra_{i,-m_i}\prod_{i<j}(m_j-m_i)t^{\sum_{i=1}^rm_i-r(r-1)/2}.\end{equation}
 As $2\g\le m_j\neq m_i\le 2\g+p-1$ and $a_{i,-m_i}\neq0$, the term of the highest degree given in \eqref{eq:8} is not zero. 
Hence $\det(W(h_1,\dots,h_r))\neq 0$. This completes the proof.
\end{proof}

\subsection{Subspace design codes}
We are now ready to present our construction of subspace design codes. 

Let $F/\F_q$ be a function field of genus $\g$ with $n+1$ pairwise distinct rational places. Let $k$ be a positive integer satisfying $1\le k\le p$, where $p$ is the characteristic of $\F_q$.  Let $\mL_k$ be the subspace defined in Equation \eqref{eq:8x}. For $s\le p$, define the encoding:
\begin{equation}
\psi:\; \mL_k\rightarrow (\F_q^s)^n,\quad f\mapsto
\begin{pmatrix}
    f(P_1) & f(P_2) & \cdots & f(P_n) \\
    f^{(1)}(P_1) & f^{(1)}(P_2) & \cdots & f^{(1)}(P_n) \\
    \vdots & \vdots & \ddots & \vdots \\
    f^{(s-1)}(P_1) & f^{(s-1)}(P_2) & \cdots & f^{(s-1)}(P_n)
\end{pmatrix}.
\end{equation}
Denote by $\mC(\PP,k)$ the image of $\psi$. 
\begin{lemma}\label{lem:4.3} If $ns>k+2\g-1$, then the code $\mC(\PP,k)$ is additive over $\F_q$ with rate $R\ge\frac{k}{ns}$ and the minimum distance at least $n-\frac {k+2\g-1}s$. 
\end{lemma}

The proof of Lemma \ref{lem:4.3} is similar to that of Lemma \ref{lem:3.1} and  we skip it.

\begin{theorem}\label{thm:4.4} Let $F/\F_q$ be a function field of genus $\g$. Assume that $\Pin$ is a rational place of $F$ and choose   $n$ pairwise distinct rational places $P_1,\dots,P_n$. Put $\PP=\{P_1,P_2,\dots,P_n\}$. If $n>\frac{k+2\g-1}s$,
then    $\mC(\PP,k)$ is a $(\tau,r)$-subspace design code with
    \[\tau=\frac s{s-r}\times\frac{k+2\g-1}{ns}.\]
\end{theorem}

\begin{proof} The proof of this theorem uses the same idea as  that of Theorem \ref{thm:3.2}. Let us give a sketch. 

 By Lemma \ref{lem:4.3}, $\psi$ is an $\F_q$-isomorphism. Let $\mA$ be an $\F_q$-subspace of  $\mC(\PP,k)$ and let $\mC_i$ be the subspace of $\mC(\PP,k)$ with the $i$th position equal to $\bo$. Let $W=\psi^{-1}(\mA)$ and $W_i=\psi^{-1}(\mA\cap \mC_i)\subseteq W$. Then $\dim_{\F_q}(W)=\dim_{\F_q}(\mA)$ and $\dim_{\F_q}(W_i)=\dim_{\F_q}(\mA\cap \mC_i)$. 

Put $r=\dim_{\F_q}(W)$ and $r_i=\dim_{\F_q}(W_i)$. 
Fix an $\F_q$-basis $\{f_1,\dots,f_r\}$ of $W$. Let $\{g_1,\dots,g_{r_i}\}$ be an $\F_q$-basis of $W_i$. Extend this basis to an $\F_q$-basis $\{g_1,\dots,g_{r}\}$ of $W$. Then there exists an invertible matrix $A_i\in\F_q^{r\times r}$ such that
\[(f_1,\dots,f_r)=(g_1,\dots,g_r)A_i.\]
By Lemma \ref{lem:2.2}, we have
\[(f_1^{(j)},\dots,f_r^{(j)})=(g_1^{(j)},\dots,g_r^{(j)})A_i\]
for all $j\ge 0$.
Combining the above equations for $j=0,1,\dots, r-1$ gives the following matrix equation 
\begin{equation}\label{eq:}
   \begin{pmatrix}
    f_1&f_2&\cdots&f_r\\
f_1^{(1)}&f_2^{(1)}&\cdots&f_r^{(1)}\\
\cdots&\cdots&\cdots&\cdots
\\
    f_1^{(r-1)}&f_2^{(r-1)}&\cdots&f_r^{(r-1)}
\end{pmatrix}=\begin{pmatrix}
    g_1&g_2&\cdots&g_r\\
g_1^{(1)}&g_2^{(1)}&\cdots&g_r^{(1)}\\
\cdots&\cdots&\cdots&\cdots
\\
    g_1^{(r-1)}&g_2^{(r-1)}&\cdots&g_r^{(r-1)}
\end{pmatrix}A_i.
\end{equation}
As $f_1,\dots,f_r$ are $\F_q$-linearly independent, the determinant $W(f_1,\dots,f_r)\neq 0$. Since every $f_i^{(j)}$ belongs to $\mL((k+2\g-1)\Pin)$, 
the determinant $W(f_1,\dots,f_r)$ belongs to the Riemann-Roch space \[\mL\left(r(k+2\g-1)\Pin\right).\] 
Note that every term of the determinant   $W(g_1,\dots,g_r)$ contains factor $\prod_{j=1}^{r_i}g_j^{(\ell_j)}$ for some $0\le \ell_1,\dots,\ell_{r_i}\le r-1$. {Moreover, $g_j^{(\ell_j)}$ has $P_i$ as a zero with multiplicity at least $s-r$ for any $1\le j\le r_i$.}
 This implies that 
\[\det(W(f_1,\dots,f_r))\in\mL\left(r(k+2\g-1)\Pin-\sum_{i=1}^n(s-r)r_iP_i\right).\]
As $\det(W(f_1,\dots,f_r))$ is not a zero function, we have
\[\deg\left(r(k+2\g-1)\Pin-\sum_{i=1}^n(s-r)r_iP_i\right)\ge 0,\]
{i.e., 
\begin{equation}\label{eq:11}r(k+2\g-1)\ge\sum_{i=1}^n(s-r)r_i.\end{equation} 
Equation \eqref{eq:11} gives }

 \[ \frac1n\sum_{i=1}^nr_i\le \frac r{s-r}\times\frac{k+2\g-1}{n}.\]
 The desired result follows.
\end{proof}
\begin{rem}{\rm 
If $\g\ll k\le p$, then we can take $s=O(1/\Ge^2)$ and $r=O(1/\Ge)$. Thus, we obtain an optimal $(R-\Ge,r)$ subspace design code. The code alphabet size is $O(n/\Ge^2)$. 
}\end{rem}

Applying Theorem \ref{thm:4.4} to function field of small genus such as projective curves, elliptic curves or hyperelliptic curves, we obtain the following result. 
\begin{cor}\label{cor:4.6}
There exists an explicit optimal $(1-R-\Ge,r)$ subspace design code over alphabet of size $O(n/\Ge^2)$.
\end{cor}

We now get a list decodable codes from our subspace design codes.

By taking $s=O(1/\Ge^2)$ and $r=O(1/\Ge)$, we obtain the following result by Lemmas \ref{lem:2.5} and \ref{lem:2.6}. 
\begin{cor}
There exists an explicit  subspace design code that is over alphabet of size is $O(n/\Ge^2)$ that is $(1-R-\Ge, O(1/\Ge))$-list decodable and has $(1-R-\Ge,O(\frac m{\Ge n}),\frac1m)$ proximity gap for any $m<\Ge n$. 
\end{cor}

\end{document}